\documentclass[prl,letterpaper,twocolumn,showpacs,superscriptaddress]{revtex4}
\usepackage{graphicx,psfrag,amsmath,amssymb,amsfonts,bbm,latexsym,color,dcolumn}

\begin{document}

\title{Quasiparticle poisoning and Josephson current fluctuations induced by
Kondo impurities}
\author{Lara Faoro}
\affiliation{Department of Physics and Astronomy, Rutgers University, 136 Frelinghuysen
Rd, Piscataway 08854, New Jersey, USA}
\affiliation{Laboratoire de Physique Th\'eorique et Hautes \'Energies, CNRS UMR 7589,
Universit\'es Paris 6 et 7, 4 place Jussieu, 75252 Paris, Cedex 05, France}
\author{Alexei Kitaev}
\affiliation{Institute for Quantum Information, California Institute of Technology,
Pasadena CA 91125, USA}
\author{Lev B. Ioffe}
\affiliation{Department of Physics and Astronomy, Rutgers University, 136 Frelinghuysen
Rd, Piscataway 08854, New Jersey, USA}
\date{\today }

\begin{abstract}
We introduce a toy model that allows us to study the physical properties of
a spin impurity coupled to the electrons in the superconducting island. We
show that when the coupling of the spin is of the order of the
superconducting gap $\Delta$ two almost degenerate subgap states are formed.
By computing the Berry phase that is associated with the superconducting
phase rotations in this model, we prove that these subgap states are
characterized by a different charge and demonstrate that the switching
between these states has the same effect as quasiparticle poisoning
(unpoisoning) of the island. We also show that an impurity coupled to both
the island and the lead generates Josepshon current fluctuations.
\end{abstract}

\maketitle




\textit{Introduction.$-$} Superconducting circuits based on small Josephson
junctions are promising candidates for the implementation of qubits \cite%
{Nakamura1999,Vion2002,Chiorescu2003,Wallraff2004,Martinis2002} and for the
development of a prototype quantum current standard \cite{Keller1999}.
Unfortunately, the performances of these devices are significantly limited
by different types of noise whose sources remain mostly unknown.
Particularly dangerous one for Single-Cooper-pair Transistors and Cooper
pair boxes is the noise produced by the incoherent tunneling of single
quasiparticles into the superconducting island. Each tunneling event changes
the island charge, thereby shifting the operation points of the device. An
important requirement for the regular operation of these devices is that
this tunneling is very rare. Despite significant experimental efforts to
reduce quasiparticle poisoning \cite%
{Aumentado2004,Yamamoto2006,Corlevi2006,Naaman2006,Palmer2007,Savin2007}  a
complete understanding of its microscopic mechanisms is still missing. 
The goal of this Letter is to show that the mechanism of the charge noise
discussed in Ref.~\cite{Faoro2005} might be also responsible for the creation of
the low energy quasiparticle traps and provides an explanation of the
puzzling features observed in quasiparticle poisoning experiments \cite%
{Court2007}. 

The work  \cite{Faoro2005} shows that Kondo-like traps located at the
Superconductor Insulator (SI) interface might produce the charge noise in
small Josephson charge qubits; similar mechanism might be responsible for
the critical current fluctuations in large superconducting contacts \cite%
{Faoro2007}. These Kondo traps are impurities with a singly occupied
electron level that carry a spin degree of freedom. Each trap is
characterized by an effective Kondo Temperature $T_{K}$ that depends
exponentially on its hybridization with the conducting electrons in the bulk
superconductor. In this mechanism both charge and critical current noise
originate from the electrons tunneling between those Kondo traps with $%
T_{K}\sim \Delta $, where $\Delta $ is the superconducting gap. In this
Letter we show that Kondo-like traps might also be responsible for
quasiparticle poisoning of the superconducting island. Further, we show that
such traps located \emph{close to} the Josephson junction generate
additional sources of critical current fluctuations due to their coupling to
the superconductors on both sides of the barrier. This mechanism for
critical current noise provides the alternative to the conventional picture
of  fluctuators blocking conducting channels in the insulating barrier. In
order to derive these results we introduce a toy model that captures the
essential physics of a spin impurity coupled to the superconducting
electrons in the superconducting island. By computing the Berry phase that
is associated with the superconducting phase rotations in this model, we
show that two different low energy states of the impurity are characterized
by a different charge. As a consequence, switching between these two low
energy states has the same effect as quasiparticle unpoisoning (poisoning)
of the island. Finally, we use this model to study the effect of the motion
of electrons between the Kondo-like traps in a Josephson junction and we
prove that if one of those traps is coupled to both the lead and the island,
these processes result in critical current fluctuations. We begin with the
review of the features of the Kondo physics that are relevant for the
following and which provides justification of the toy model.

The behavior of a spin-$1/2$ impurity coupled antiferromagnetically with an
exchange constant $J$ to an electron gas characterized by a constant density
of states $\rho _{0}$ within a bandwidth $D$ is completely different at high
and low temperature regimes. In the former, the electrons scatter off the
impurity inelastically in a spin-flip process while at low temperatures the
impurity is screened by the electrons forming a bound singlet state leaving
only elastic scattering. The crossover takes place at energy scale of Kondo
temperature ${T_{K}\sim De^{-1/J\rho _{0}}}$. All relevant physics is
described by the Anderson Hamiltonian: ${H=H_{lead}+H_{d}+H_{sd}}$, where ${%
H_{lead}=\epsilon _{\kappa }c_{\kappa ,\sigma }^{\dagger }c_{\kappa ,\sigma }%
}$ and 
\begin{equation}
\begin{split}
H_{d}& =\sum_{\sigma }\epsilon _{d}c_{d\sigma }^{\dagger }c_{d\sigma
}+Un_{d\uparrow }n_{d\downarrow }\; \\
H_{sd}& =\sum_{\kappa \sigma }\left( Vc_{\kappa \sigma }^{\dagger
}c_{d\sigma }+h.c.\right) \;
\end{split}
\label{Anderson}
\end{equation}%
Here $c_{\kappa \sigma }$ $(c_{\kappa \sigma }^{\dagger })$ is the
annihilation (creation) operator for an electron in the band, $c_{d\sigma }$ 
$(c_{d\sigma }^{\dagger })$ is the annihilation (creation) operator for an
electron on the impurity site. $U$ denotes the Coulomb onsite repulsion of
the trap and ${n_{d\sigma }=c_{d\sigma }^{\dagger }c_{d\sigma }}$. In the
limit of strong onsite repulsion, i.e. ${U\gg \Gamma }$, where ${\Gamma =\pi
\rho _{0}V^{2}}$ defines the hybridization to the conducting electrons,
using the Schrieffer-Wolff transformation \cite{Schrieffer1966}, we can map
the Hamiltonian given in Eq.(\ref{Anderson}) to the Kondo model \cite{Kondo}%
: 
\begin{equation}
H=H_{lead}+\sum_{\kappa \kappa ^{\prime }}J_{\kappa \kappa ^{\prime }}\vec{S}%
\cdot c_{\kappa \sigma _{1}}^{\dagger }\vec{\sigma}_{\sigma _{1}\sigma
_{2}}c_{\kappa ^{\prime }\sigma _{2}}  \label{Kondo}
\end{equation}%
Here ${J_{\kappa \kappa ^{\prime }}=J=8V^{2}/U}$. This mapping neglects the
particle hole asymmetry of the original problem, the effects that are small
in $T_{K}/D$ but might have important physical consequences. Notice that in
the limit ${J/D\rightarrow 1}$, the Kondo temperature becomes ${T_{K}\approx
J}$.

A more complicated problem is presented by the impurity interacting with the
conduction electrons in the superconductor. At present, only qualitative
arguments and numerical results are available which show that the
competition between the Kondo temperature of the trap and the
superconducting gap $\Delta $ of the lead results in three different regimes
for the system (impurity+superconductor): (i) ${T_{K}\ll \Delta }$, where
the ground state of the system is a doublet and it is characterized by an
odd number of electrons; (ii) ${T_{K}\gg \Delta }$ where the ground state of
the system is a singlet, the electron of the impurity forms a bound state
with the superconducting electrons and the total number of electrons is
even; (iii) ${T_{K}\sim \Delta }$ where the singlet and doublet states
become almost degenerate.

\textit{Toy model and quasiparticle poisoning.$-$} In order to formulate a
simplified model that captures the effects of an impurity interacting with
the superconducting electrons, we notice that the Kondo physics can be
viewed as a result of the 'poor-man scaling' in which the high energy
degrees of freedom are gradually integrated out resulting in the logarithmic
growth of the effective interaction. In the presence of the superconducting
gap the process of integration has to stop at ${\Delta }$. If at this moment
the interaction is comparable with ${\Delta ,}$ a bound subgap state can be
formed in agreement with the results described above. This shows that the
essential physics can be captured by a simple model in which the spin
interacts with a single electron mode with the coupling constant $J\sim {%
\Delta }$ and energy ${\varepsilon <\Delta }$ that is described by the BCS
Hamiltonian: 
\begin{equation*}
H_{BCS}=\varepsilon \sum_{i=k\uparrow ,-k\downarrow }c_{i}^{\dagger
}c_{i}+\Delta \left( e^{i\theta }c_{k\uparrow }^{\dagger }c_{-k\downarrow
}^{\dagger }+e^{-i\theta }c_{k\uparrow }c_{-k\downarrow }\right) .
\end{equation*}%
The interaction between the impurity and the superconducting electrons is
described by the spin exchange coupling given in Eq.(\ref{Kondo}): 
\begin{equation}
H_{toy}=H_{BCS}+\sum_{k,k^{\prime }}J_{k,k^{\prime }}\vec{S}\cdot c_{\kappa
\sigma _{1}}^{\dagger }\vec{\sigma}_{\sigma _{1}\sigma _{2}}c_{\kappa
^{\prime }\sigma _{2}}\;  \label{model1}
\end{equation}%
We assume that the coupling is isotropic: i.e. ${J_{k,k^{\prime }}\equiv
J\approx T_{K}\sim \Delta }$. The Hamiltonian (\ref{model1}) can be readily
diagonalized. We choose the state basis: ${\{|i\rangle _{k\uparrow
},|j\rangle _{-k\downarrow },|\sigma \rangle _{imp}\}}$ where ${i,j=0,1}$
denotes respectively the absence or presence of the quasiparticle in the
single electron mode while ${\sigma =\Uparrow ,\Downarrow }$ represents the
spin configuration up or down of the electron in the trap and find the
lowest eigenvalues: 
\begin{equation}
\begin{split}
E_{0}& =\varepsilon -\sqrt{\Delta ^{2}+\varepsilon ^{2}}\; \\
E_{1}& =\varepsilon -\frac{3}{2}T_{K}\;
\end{split}%
\end{equation}%
corresponding respectively to the (non normalized) doublet and singlet
states: 
\begin{equation}
\begin{split}
& |D_{\sigma }\rangle =\left[ -\frac{(\varepsilon +\sqrt{\Delta
^{2}+\varepsilon ^{2})}}{|\Delta |}e^{-i\theta }|00\rangle +|11\rangle %
\right] |\sigma \rangle \; \\
& |S\rangle =-|01\Downarrow \rangle +|10\Uparrow \rangle \;
\end{split}%
\end{equation}%
As expected, the spin impurity interacting with the conduction electrons in
the superconductor leads to the formation of weak Kondo subgap states.
Notice that the subgap states have different properties: the doublet is
characterized by an odd number of electrons, its degeneracy is due to the
spin degree of freedom of the trap while the singlet state is a maximally
entangled state with even number of electrons. Depending on the ratio ${%
T_{K}/\Delta }$, the ground state of the system can be either doublet or
singlet. At a special value $\displaystyle{T_{K}^{\ast }=\frac{2}{3}\sqrt{%
\varepsilon ^{2}+\Delta ^{2}}}$ singlet and doublet states are degenerate
while for traps with ${T_{K}\approx T_{K}^{\ast }\sim \Delta }$ singlet and
doublet states are almost degenerate.

So far, we have reproduced in a simplified manner the main content of the
numerical results. We want now to show that singlet and doublet states
differ in another important feature: namely, the doublet corresponds to a
non zero off-set charge in the superconducting island. For this purpose, we
recall that the operator $n$, describing the excess number of Cooper pairs
on the island and the operator $\theta $, representing the superconducting
phase, are conjugate variables, i.e. ${\left[ \theta ,n\right] =i}$, and
that the Hamiltonian of the island/box $H_{isl/box}$ is invariant with
respect to the local gauge transformation: ${U^{-1}H_{isl/box}U}$ where ${%
U=e^{in_{g}\theta }}$ and $n_{g}$ is the off-set charge induced in the
island in unit $2e$, that plays a role similar to the vector potential
appearing in the Hamiltonian of an electron in a magnetic field. One
consequence of these observations, is that we can deduce the value of the
off-set charge induced by the Kondo impurity in the singlet and doublet
states from the value of the Berry phases associated to these states. We
find that 
\begin{eqnarray}
\oint n\cdot d\theta & =&i\int_{0}^{2\pi }\langle S|\frac{\partial }{\partial
\theta }|S\rangle =0\; \notag \\ 
\oint n\cdot d\theta & =& i\int_{0}^{2\pi }\langle D_{\sigma }|\frac{\partial 
}{\partial \theta }|D_{\sigma }\rangle =\pi \left( 1+\frac{\varepsilon }{%
\sqrt{\varepsilon ^{2}+\Delta ^{2}}}\right) \,\;
\notag
\end{eqnarray}%
Notice that a Kondo impurity induces a non-zero off-set charge in the
superconducting island \emph{only} when the system (impurity+superconductor)
is in the doublet state, while in the singlet state the electron charge is
fully screened. Moreover, when ${\varepsilon \rightarrow 0}$, we find that $%
n\rightarrow \frac{1}{2}$, i.e. exactly one electron is induced on the
superconducting island. Let us now consider two Kondo-traps with ${T_{K}\sim
\Delta }$ located at the SI interfaces, one on the superconducting island,
another on the lead within distance $\xi $ from each other. The electron
tunneling process across the junction couples these traps. When the Kondo
trap located at the SI interface in the island switches between singlet and
doublet, the parity of the island changes from even to odd. For this process
to be physically relevant, the energy difference between these states should
be smaller than $T$. Thus, the pairs of Kondo subgap states with close
energy levels might be responsible for the quasiparticle poisoning in
superconducting devices. Notice that if the two traps are located on the
same side of the barrier, the switching between singlet and doublet caused
by the tunneling of quasiparticles through the superconductor results in
charge fluctuations, i.e. the entire process can be viewed as a new type of
a charge fluctuator.

We now discuss the implications for the recent experiments where
quasiparticle tunneling rates were measured with microsecond resolution \cite%
{Naaman2006,Ferguson2006,Court2007} in a micrometer-sized island with
capacitive gate electrode that was probed by two Josephson junctions. The
island charging energy is modulated by the gate as ${%
E_{c}^{n}(n_{g})=E_{c}(n-n_{g})^{2}}$, where $E_{c}=e^{2}/2C$, $e$ is the
electron charge and $C$ is the total island capacitance, ${n_{g}=C_{g}V_{g}/e%
}$ is the normalized gate charge and $n$ is the integer number of excess
charges on the island. At ${n_{g}=1}$ the electrostatic energy of the system
is minimized when unpaired electrons reside in the superconducting island.
Thus, at ${n_{g}=1}$ the island is a trap for a quasiparticle with depth $%
\displaystyle{\delta E=E_{c}-E_{J}/2+\Delta _{l}-\Delta _{i}}$. Here $\Delta
_{i}$, $\Delta _{l}$ are the superconducting gap of the island, lead and $%
E_{J}$ is the Josephson energy. A model suggested by Aumentado et al. \cite%
{Aumentado2004} explains many features of quasiparticle poisoning of the
island. In this model some unknown non equilibrium source of quasiparticles
produces them in the leads. Quasiparticles are able to tunnel onto the
island which acts as a trap. Subsequently the trapped quasiparticle is
thermally excited (unpoisoning) out of the trap and the island returns to
its even state. Relevant implications of this model are that the
quasiparticle poisoning can be reduced by putting normal metal leads (QP
traps) close to the junctions in order to filter the quasiparticles and by
making ${\Delta _{i}>\Delta _{l}}$, because it works as a barrier, which
prevents non equilibrium quasiparticles in the leads from entering the
island. Experiments showed that these ideas help to reduce quasiparticle
poisoning, but do not eliminate it. In particular, the effect of
quasiparticle traps have been recently studied in Ref. \cite{Court2007}. In
these experiments, two similar Cooper pair boxes were fabricated with (QT)
or without (NT) quasiparticles traps attached to the leads. The island was
biased at ${n_{g}=1}$ and the dynamics of the quasiparticles captured by the
island was characterized by their incoming ($t_{even}$) and outgoing ($%
t_{odd}$) rates. One expects that the incoming process involves
quasiparticle tunneling into the island and its relaxation to the bottom of
the well, while the reverse process involves thermal excitation. As a
result, the outgoing rate should be smaller by a factor ${\propto e^{-\delta
E/KT}}$ than the incoming rate. This is in contrast with the data that show
that the trapping and escape rates are roughly equal and temperature
independent below ${T\lesssim 200}$mK. However, their values are
dramatically different in the devices with or without traps: ${t_{even}\sim
t_{odd}\approx (10^{2}-10^{3})\mu s}$ (QT) and ${t_{even}\sim t_{odd}\approx
(0.1-1)\mu s}$ (NT).

\begin{figure}[h]
\centering
\includegraphics[scale=0.6]{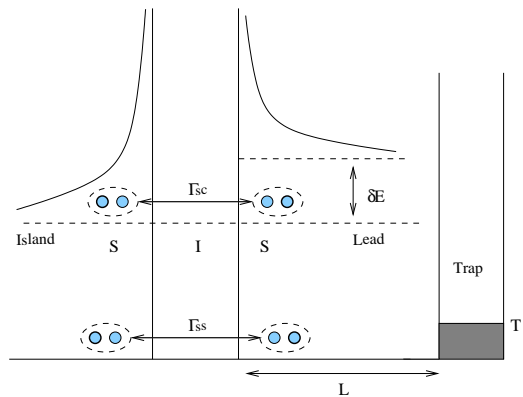}
\caption{Sketches of quasiparticle poisoning and unpoisoning due to
quasiparticle tunneling between weak Kondo subgap states located at the
lead/island SI interfaces.}
\label{fig1}
\end{figure}

The presence of subgap states in the lead and the island provides a
different scenario where two new processes are present: quasiparticles with
energies above ${\Delta _{i}-E_{c}+E_{J}/2}$ tunnel from a subgap state in
the lead to the continuum in the island while quasiparticles below these
energies tunnel between the subgap states in the island and in the lead (see
Fig.~\ref{fig1}). The rate of the former process is ${\Gamma _{sc}=G\delta }$%
, where ${G\sim 1}$ is the conductance of the barrier in the units of $%
e^{2}/\hbar $ and ${\delta =1/V_{i}\nu _{Al}}$ is the typical level spacing
in the island. For a typical island of volume ${V_{i}=750\text{nm}\times 125%
\text{nm}\times 7\text{nm}}$ and a typical Al electron density of states ${%
\nu _{Al}\sim 35/eVnm^{3}}$, we estimate ${\Gamma _{sc}\approx 10^{-7}\text{s%
}^{-1}}$. The rate of the exchange process between subgap states is much slower ${\Gamma _{ss}\ll
\Gamma _{sc}}$ because it occurs between two localized states and it depends
on the level width of the state. In both cases the tunneling does not
involve a significant energy transfer, so we expect these rates to be
temperature independent. In the presence of quasiparticles with energy
larger than ${\Delta _{i}-E_{c}+E_{J}/2}$ the first process dominates and the
observable rate is ${\Gamma _{sc}}$. The presence of quasiparticle traps
attached to the leads eliminate high energy quasiparticles and the rate
decreases to ${\Gamma _{ss}}$ in agreement with observations. This scenario
can be checked by fabricating QT devices with traps in the lead located at
different distance $L\sim \xi $. The presence of these traps at sufficiently
small distance broadens the levels of the subgap states in the leads and
consequently it should increase the tunneling rate into the island as ${%
\propto e^{-L/\xi }}$, where $\xi $ is the coherence length of the
superconductor. The estimates of the rate assume that only a few subgap
states are active at the same time, large number of these states would make
the effective rate higher; we do not know the density of these states and
their occupation in a realistic system.

\textit{Josephson current fluctuations.$-$} Kondo-like traps with ${%
T_{K}\sim \Delta }$ provide additional source of the critical current
fluctuations when they are located close to the Josephson junction barrier
because in this case the spin of the Kondo trap is coupled both to the
electrons in the island and in the lead. This can be easily seen by
including in our toy model the electrons of the lead and a tunneling between
the lead and the island. The Hamiltonian becomes ${%
H=H_{toy}^{(1)}+H_{BCS}^{(2)}+H_{T}^{qp}}$ and the quasiparticle tunneling
through the junction barrier is given by: 
\begin{equation}
H_{T}^{qp}=|T_{k_{1},k_{2}}|\left[ c_{k_{1}\uparrow }^{\dagger
}c_{k_{2}\uparrow }+c_{-k_{1}\downarrow }^{\dagger }c_{-k_{2}\downarrow
}+h.c.\right] \;  \notag
\end{equation}%
We assume that the superconducting leads are equal and we calculate the
correction to the lowest energy eigenvalues and eigenvectors at the second
order in perturbation theory in the tunneling ${|T_{k_{1},k_{2}}|\approx {%
\mathcal{T}}}$. We find that the correction to the singlet state depends on
the phase difference ${\varphi =\theta _{1}-\theta _{2}}$. The dependence on 
${\varphi }$ implies additional contribution to the Josephson current. A
straightforward but lengthy calculation, gives the contribution of the Kondo
impurity in the barrier to the Josephson current: 
\begin{equation}
\delta I_{c}\approx \frac{{\mathcal{T}}^{2}}{J}\frac{\Delta ^{2}}{%
\varepsilon ^{2}+\Delta ^{2}}\sin \varphi   \label{flucri}
\end{equation}%
To find the parameters of the toy model that describe the physical situation
in which a many channel junction couples a small metallic island to a large
superconducting lead, we compare the pairing field induced on the state $%
k_{1}$ in the islands by the superconducting order in the leads in the
realistic situation ($\nu {\mathcal{T}}^{2}\sim G\delta )$ with the
corresponding quantity in the simplified model (${\mathcal{T}}^{2}/\Delta $)
and get $\displaystyle{\frac{{\mathcal{T}}^{2}}{J}=G\delta }$. Notice that
this reasoning holds for small superconducting islands whose size is less
than the superconducting correlation length $\xi $. For larger islands the
impurities at a distance larger than $\xi $ from the junction are coupled
exponentially weakly to the superconductor on the other side of the barrier.

\textit{Conclusions.$-$} We have shown that subgap states generated by
magnetic impurities due to the competition between superconducting pairing
and Kondo effect act as very efficient quasiparticle traps. We argued that
the presence of such states in a typical Single-Cooper-pair Transistor and
Cooper pair box might explain the results of recent experiments where
unexpected poisoning/unpoisoning rates were observed. We have also shown the
same subgap states generate critical current noise.

This work was supported by the National Security Agency (NSA) under Army
Research Office (ARO) contract number W911NF-06-1-0208 and NSF ECS 0608842.

\vspace*{-2mm} 
\bibliography{QP_poisoningF.bbl}

\begin{thebibliography}{18}
\expandafter\ifx\csname natexlab\endcsname\relax\def\natexlab#1{#1}\fi
\expandafter\ifx\csname bibnamefont\endcsname\relax
  \def\bibnamefont#1{#1}\fi
\expandafter\ifx\csname bibfnamefont\endcsname\relax
  \def\bibfnamefont#1{#1}\fi
\expandafter\ifx\csname citenamefont\endcsname\relax
  \def\citenamefont#1{#1}\fi
\expandafter\ifx\csname url\endcsname\relax
  \def\url#1{\texttt{#1}}\fi
\expandafter\ifx\csname urlprefix\endcsname\relax\def\urlprefix{URL }\fi
\providecommand{\bibinfo}[2]{#2}
\providecommand{\eprint}[2][]{\url{#2}}

\bibitem[{\citenamefont{Nakamura et~al.}(1999)\citenamefont{Nakamura, Pashkin,
  and Tsai}}]{Nakamura1999}
\bibinfo{author}{\bibfnamefont{Y.}~\bibnamefont{Nakamura}},
  \bibinfo{author}{\bibfnamefont{Y.~A.} \bibnamefont{Pashkin}},
  \bibnamefont{and} \bibinfo{author}{\bibfnamefont{J.~S.} \bibnamefont{Tsai}},
  \bibinfo{journal}{Nature (London)} \textbf{\bibinfo{volume}{398}},
  \bibinfo{pages}{786} (\bibinfo{year}{1999}).

\bibitem[{\citenamefont{Vion et~al.}(2002)\citenamefont{Vion, Aassime, Cottet,
  Joyez, Pothier, Urbina, Esteve, and Devoret}}]{Vion2002}
\bibinfo{author}{\bibfnamefont{D.}~\bibnamefont{Vion}},
  \bibinfo{author}{\bibfnamefont{A.}~\bibnamefont{Aassime}},
  \bibinfo{author}{\bibfnamefont{A.}~\bibnamefont{Cottet}},
  \bibinfo{author}{\bibfnamefont{P.}~\bibnamefont{Joyez}},
  \bibinfo{author}{\bibfnamefont{H.}~\bibnamefont{Pothier}},
  \bibinfo{author}{\bibfnamefont{C.}~\bibnamefont{Urbina}},
  \bibinfo{author}{\bibfnamefont{D.}~\bibnamefont{Esteve}}, \bibnamefont{and}
  \bibinfo{author}{\bibfnamefont{M.~H.} \bibnamefont{Devoret}},
  \bibinfo{journal}{Science} \textbf{\bibinfo{volume}{296}},
  \bibinfo{pages}{886} (\bibinfo{year}{2002}).

\bibitem[{\citenamefont{Chiorescu et~al.}(2003)\citenamefont{Chiorescu,
  Nakamura, Harmans, and Mooij}}]{Chiorescu2003}
\bibinfo{author}{\bibfnamefont{I.}~\bibnamefont{Chiorescu}},
  \bibinfo{author}{\bibfnamefont{Y.}~\bibnamefont{Nakamura}},
  \bibinfo{author}{\bibfnamefont{C.~J. P.~M.} \bibnamefont{Harmans}},
  \bibnamefont{and} \bibinfo{author}{\bibfnamefont{J.~E.} \bibnamefont{Mooij}},
  \bibinfo{journal}{Science} \textbf{\bibinfo{volume}{299}},
  \bibinfo{pages}{1869} (\bibinfo{year}{2003}).

\bibitem[{\citenamefont{Wallraff et~al.}(2004)\citenamefont{Wallraff, Schuster,
  Blais, Frunzio, Huang, Majer, Kumar, Girvin, and Schoelkopf}}]{Wallraff2004}
\bibinfo{author}{\bibfnamefont{A.}~\bibnamefont{Wallraff}},
  \bibinfo{author}{\bibfnamefont{D.~I.} \bibnamefont{Schuster}},
  \bibinfo{author}{\bibfnamefont{A.}~\bibnamefont{Blais}},
  \bibinfo{author}{\bibfnamefont{L.}~\bibnamefont{Frunzio}},
  \bibinfo{author}{\bibfnamefont{R.~S.} \bibnamefont{Huang}},
  \bibinfo{author}{\bibfnamefont{J.}~\bibnamefont{Majer}},
  \bibinfo{author}{\bibfnamefont{S.}~\bibnamefont{Kumar}},
  \bibinfo{author}{\bibfnamefont{S.~M.} \bibnamefont{Girvin}},
  \bibnamefont{and} \bibinfo{author}{\bibfnamefont{R.~J.}
  \bibnamefont{Schoelkopf}}, \bibinfo{journal}{Nature}
  \textbf{\bibinfo{volume}{432}}, \bibinfo{pages}{162} (\bibinfo{year}{2004}).

\bibitem[{\citenamefont{Martinis et~al.}(2002)\citenamefont{Martinis, Nam,
  Aumentado, and Urbina}}]{Martinis2002}
\bibinfo{author}{\bibfnamefont{J.~M.} \bibnamefont{Martinis}},
  \bibinfo{author}{\bibfnamefont{S.}~\bibnamefont{Nam}},
  \bibinfo{author}{\bibfnamefont{J.}~\bibnamefont{Aumentado}},
  \bibnamefont{and} \bibinfo{author}{\bibfnamefont{C.}~\bibnamefont{Urbina}},
  \bibinfo{journal}{Phys. Rev. Lett.} \textbf{\bibinfo{volume}{89}},
  \bibinfo{pages}{117901} (\bibinfo{year}{2002}).

\bibitem[{\citenamefont{Keller et~al.}(1999)\citenamefont{Keller, Eichenberger,
  Martinis, and Zimmerman}}]{Keller1999}
\bibinfo{author}{\bibfnamefont{M.~W.} \bibnamefont{Keller}},
  \bibinfo{author}{\bibfnamefont{A.~L.} \bibnamefont{Eichenberger}},
  \bibinfo{author}{\bibfnamefont{J.~M.} \bibnamefont{Martinis}},
  \bibnamefont{and} \bibinfo{author}{\bibfnamefont{N.~M.}
  \bibnamefont{Zimmerman}}, \bibinfo{journal}{Science}
  \textbf{\bibinfo{volume}{285}}, \bibinfo{pages}{1706} (\bibinfo{year}{1999}).

\bibitem[{\citenamefont{Aumentado et~al.}(2004)\citenamefont{Aumentado, Keller,
  Martinis, and Devoret}}]{Aumentado2004}
\bibinfo{author}{\bibfnamefont{J.}~\bibnamefont{Aumentado}},
  \bibinfo{author}{\bibfnamefont{M.~W.} \bibnamefont{Keller}},
  \bibinfo{author}{\bibfnamefont{J.~M.} \bibnamefont{Martinis}},
  \bibnamefont{and} \bibinfo{author}{\bibfnamefont{M.~H.}
  \bibnamefont{Devoret}}, \bibinfo{journal}{Phys. Rev. Lett.}
  \textbf{\bibinfo{volume}{92}}, \bibinfo{eid}{066802} (\bibinfo{year}{2004}).

\bibitem[{\citenamefont{Yamamoto et~al.}(2006)\citenamefont{Yamamoto, Nakamura,
  Pashkin, Astafiev, and Tsai}}]{Yamamoto2006}
\bibinfo{author}{\bibfnamefont{T.}~\bibnamefont{Yamamoto}},
  \bibinfo{author}{\bibfnamefont{Y.}~\bibnamefont{Nakamura}},
  \bibinfo{author}{\bibfnamefont{Y.~A.} \bibnamefont{Pashkin}},
  \bibinfo{author}{\bibfnamefont{O.}~\bibnamefont{Astafiev}}, \bibnamefont{and}
  \bibinfo{author}{\bibfnamefont{J.~S.} \bibnamefont{Tsai}},
  \bibinfo{journal}{Appl. Phys. Lett.} \textbf{\bibinfo{volume}{88}},
  \bibinfo{pages}{212509} (\bibinfo{year}{2006}).

\bibitem[{\citenamefont{Corlevi et~al.}(2006)\citenamefont{Corlevi, Guichard,
  Hekking, and Haviland}}]{Corlevi2006}
\bibinfo{author}{\bibfnamefont{S.}~\bibnamefont{Corlevi}},
  \bibinfo{author}{\bibfnamefont{W.}~\bibnamefont{Guichard}},
  \bibinfo{author}{\bibfnamefont{F.~W.~J.} \bibnamefont{Hekking}},
  \bibnamefont{and} \bibinfo{author}{\bibfnamefont{D.~B.}
  \bibnamefont{Haviland}}, \bibinfo{journal}{Phys. Rev. B}
  \textbf{\bibinfo{volume}{74}}, \bibinfo{eid}{224505} (\bibinfo{year}{2006}).

\bibitem[{\citenamefont{Naaman and Aumentado}(2006)}]{Naaman2006}
\bibinfo{author}{\bibfnamefont{O.}~\bibnamefont{Naaman}} \bibnamefont{and}
  \bibinfo{author}{\bibfnamefont{J.}~\bibnamefont{Aumentado}},
  \bibinfo{journal}{Phys. Rev. B} \textbf{\bibinfo{volume}{73}},
  \bibinfo{eid}{172504} (\bibinfo{year}{2006}).

\bibitem[{\citenamefont{Palmer et~al.}(2007)\citenamefont{Palmer, Sanchez,
  Naik, Manheimer, Schneiderman, Echternach, and Wellstood}}]{Palmer2007}
\bibinfo{author}{\bibfnamefont{B.~S.} \bibnamefont{Palmer}},
  \bibinfo{author}{\bibfnamefont{C.~A.} \bibnamefont{Sanchez}},
  \bibinfo{author}{\bibfnamefont{A.}~\bibnamefont{Naik}},
  \bibinfo{author}{\bibfnamefont{M.~A.} \bibnamefont{Manheimer}},
  \bibinfo{author}{\bibfnamefont{J.~F.} \bibnamefont{Schneiderman}},
  \bibinfo{author}{\bibfnamefont{P.~M.} \bibnamefont{Echternach}},
  \bibnamefont{and} \bibinfo{author}{\bibfnamefont{F.~C.}
  \bibnamefont{Wellstood}}, \bibinfo{journal}{Phys. Rev. B}
  \textbf{\bibinfo{volume}{76}}, \bibinfo{eid}{054501} (\bibinfo{year}{2007}).

\bibitem[{\citenamefont{Savin et~al.}(2007)\citenamefont{Savin, Meschke,
  Pekola, Pashkin, Li, Im, and Tsai}}]{Savin2007}
\bibinfo{author}{\bibfnamefont{A.~M.} \bibnamefont{Savin}},
  \bibinfo{author}{\bibfnamefont{M.}~\bibnamefont{Meschke}},
  \bibinfo{author}{\bibfnamefont{J.~P.} \bibnamefont{Pekola}},
  \bibinfo{author}{\bibfnamefont{Y.~A.} \bibnamefont{Pashkin}},
  \bibinfo{author}{\bibfnamefont{T.~F.} \bibnamefont{Li}},
  \bibinfo{author}{\bibfnamefont{H.}~\bibnamefont{Im}}, \bibnamefont{and}
  \bibinfo{author}{\bibfnamefont{J.~S.} \bibnamefont{Tsai}},
  \bibinfo{journal}{Appl. Phys. Lett..} \textbf{\bibinfo{volume}{91}},
  \bibinfo{pages}{63512} (\bibinfo{year}{2007}).

\bibitem[{\citenamefont{Faoro and Ioffe}(2006)}]{Faoro2005}
\bibinfo{author}{\bibfnamefont{L.}~\bibnamefont{Faoro}} \bibnamefont{and}
  \bibinfo{author}{\bibfnamefont{L.~B.} \bibnamefont{Ioffe}},
  \bibinfo{journal}{Phys. Rev. Lett} \textbf{\bibinfo{volume}{94}},
  \bibinfo{pages}{047001} (\bibinfo{year}{2006}).

\bibitem[{\citenamefont{Court et~al.}(2007)\citenamefont{Court, Ferguson,
  Lutchyn, and Clark}}]{Court2007}
\bibinfo{author}{\bibfnamefont{N.~A.} \bibnamefont{Court}},
  \bibinfo{author}{\bibfnamefont{A.~J.} \bibnamefont{Ferguson}},
  \bibinfo{author}{\bibfnamefont{R.}~\bibnamefont{Lutchyn}}, \bibnamefont{and}
  \bibinfo{author}{\bibfnamefont{R.~G.} \bibnamefont{Clark}}
  (\bibinfo{year}{2007}), \bibinfo{note}{arXiv:0710.2760}.

\bibitem[{\citenamefont{Faoro and Ioffe}(2007)}]{Faoro2007}
\bibinfo{author}{\bibfnamefont{L.}~\bibnamefont{Faoro}} \bibnamefont{and}
  \bibinfo{author}{\bibfnamefont{L.~B.} \bibnamefont{Ioffe}},
  \bibinfo{journal}{Phys. Rev. B} \textbf{\bibinfo{volume}{75}},
  \bibinfo{eid}{132505} (\bibinfo{year}{2007}).

\bibitem[{\citenamefont{Schrieffer and Wolff}(1966)}]{Schrieffer1966}
\bibinfo{author}{\bibfnamefont{J.~R.} \bibnamefont{Schrieffer}}
  \bibnamefont{and} \bibinfo{author}{\bibfnamefont{P.~A.} \bibnamefont{Wolff}},
  \bibinfo{journal}{Phys. Rev.} \textbf{\bibinfo{volume}{149}},
  \bibinfo{pages}{491} (\bibinfo{year}{1966}).

\bibitem[{\citenamefont{Kondo}(1964)}]{Kondo}
\bibinfo{author}{\bibfnamefont{J.}~\bibnamefont{Kondo}},
  \bibinfo{journal}{Prog. Theor. Phys.} \textbf{\bibinfo{volume}{32}},
  \bibinfo{pages}{37} (\bibinfo{year}{1964}).

\bibitem[{\citenamefont{Ferguson et~al.}(2006)\citenamefont{Ferguson, Court,
  Hudson, and Clark}}]{Ferguson2006}
\bibinfo{author}{\bibfnamefont{A.~J.} \bibnamefont{Ferguson}},
  \bibinfo{author}{\bibfnamefont{N.~A.} \bibnamefont{Court}},
  \bibinfo{author}{\bibfnamefont{F.~E.} \bibnamefont{Hudson}},
  \bibnamefont{and} \bibinfo{author}{\bibfnamefont{R.~G.} \bibnamefont{Clark}},
  \bibinfo{journal}{Phys. Rev. Lett.} \textbf{\bibinfo{volume}{97}},
  \bibinfo{eid}{106603} (\bibinfo{year}{2006}).

\end{thebibliography}

\end{document}